\documentclass[10pt,journal,compsoc]{IEEEtran}

\ifCLASSOPTIONcompsoc
  \usepackage[nocompress]{cite}
\else
  \usepackage{cite}
\fi

\usepackage[ruled,vlined,linesnumbered]{algorithm2e}
\usepackage[lambda,advantage,operators,sets,adversary,landau,probability,notions,logic,ff,mm,primitives,events,complexity,asymptotics,keys]{cryptocode}
\usepackage{booktabs}
\usepackage{threeparttable}
\usepackage{multirow}
\usepackage{amsthm}

\theoremstyle{definition}
\newtheorem{definition}{Definition}[section]
\newtheorem{theorem}{Theorem}[section]

\begin{document}

\title{Blockchain-based Transparency Framework for Privacy Preserving Third-party Services}

\author{
    Runhua~Xu,~\IEEEmembership{Member,~IEEE,}
    Chao~Li,~\IEEEmembership{Member,~IEEE,}
    and~James~Joshi,~\IEEEmembership{Senior~Member,~IEEE}
    \IEEEcompsocitemizethanks{
        \IEEEcompsocthanksitem Runhua Xu is with the IBM Research, San Jose, CA, United States, 95120. 
        Chao Li is with Beijing Key Laboratory of Security and Privacy in Intelligent Transportation, Beijing Jiaotong University, Beijing, China, 100044. 
        Runhua Xu and Chao Li are corresponding authors.
        James Joshi is with School of Computing and Information, University of Pittsburgh, Pittsburgh, PA, United States, 15260.\protect\\
        E-mail:runhua@ibm.com, li.chao@bjtu.edu.cn, jjoshi@pitt.edu
        }
}

\IEEEtitleabstractindextext{%

\begin{abstract}
Increasingly, information systems rely on computational, storage, and network resources deployed in third-party facilities such as cloud centers and edge nodes. Such an approach further exacerbates cybersecurity concerns constantly raised by numerous incidents of security and privacy attacks resulting in data leakage and identity theft, among others. These have, in turn, forced the creation of stricter security and privacy-related regulations and have eroded the trust in cyberspace. In particular, security-related services and infrastructures, such as Certificate Authorities (CAs) that provide digital certificate services and Third-Party Authorities (TPAs) that provide cryptographic key services, are critical components for establishing trust in crypto-based privacy-preserving applications and services. To address such trust issues, various \textit{transparency} frameworks and approaches have been recently proposed in the literature.
This paper proposes \texttt{TAB} framework that provides transparency and trustworthiness of third-party authority and third-party facilities using blockchain techniques for emerging crypto-based privacy-preserving applications.
\texttt{TAB} employs the Ethereum blockchain as the underlying public ledger and also includes a novel smart contract to automate accountability with an incentive mechanism that motivates users to participate in auditing, and punishes unintentional or malicious behaviors. 
We implement \texttt{TAB} and show through experimental evaluation in the Ethereum official test network, Rinkeby, that the framework is efficient.
We also formally show the security guarantee provided by \texttt{TAB}, and analyze the privacy guarantee and trustworthiness it provides.
\end{abstract}

\begin{IEEEkeywords}
Transparency, Trustworthiness, Third-party Authority, Blockchain, Ethereum, Smart Contract, Functional Encryption
\end{IEEEkeywords}
}

\maketitle

\IEEEdisplaynontitleabstractindextext

\IEEEpeerreviewmaketitle

\section{Introduction}
\begin{figure}[!t]
    \centering
    \includegraphics[width=0.3\textwidth]{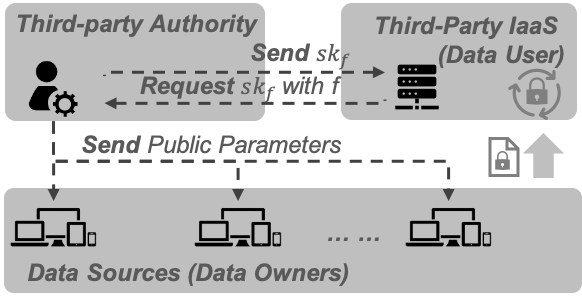}
    \caption{Illustration of privacy-preserving third-party service enabled system.}
    \label{fig:application}
\end{figure}

\IEEEPARstart{I}{ncreasingly}, information systems are being built on the third-party facilities or use external services. 
This is beneficial to many enterprises as it lowers costs and allows them to keep their focus on business missions. 
On the other hand, increasing cybersecurity incidents such as cybersecurity attacks including those leading to data leakage and identity theft are amplifying users' concerns with regards to their sensitive personal data that is collected, stored, and processed on the third-party facilities.
Furthermore, regulations such as General Data Protection Regulation (GDPR) and California Consumer Privacy Act (CCPA) introduce stricter compliance requirements for enterprise information systems.
\figurename~\ref{fig:application} illustrates the architecture of a typical privacy-preserving third-party service enabled system, as illustrated in a variety of existing work \cite{sans2018reading, xu2019cryptonn, xu2019hybridalpha, ryffel2019partially}, where the personal data is protected by a cryptosystem, and the encrypted data is collected and processed by a third-party IaaS, while the public key and private key services are provided by the third-party authority (TPA). 
Usually, the third-party entities are assumed to be \textit{honest-but-curious}, and a TPA is typically \textit{fully trusted}.

To address the trust and compliance issues on these service providers, especially, the security-related service providers such as certificate authorities (CAs), key directories (KDs) and TPAs that provide services of certification, public key lookup and private key generation in existing cryptographic key infrastructures, various \textit{transparency} approaches that provision \textit{openness} and \textit{accountability} have been recently proposed \cite{laurie2013certificate,laurie2014certificate,melara2015coniks,chase2016transparency,bonneau2016ethiks,chase2019seemless} to increase users' trust or confidence in such cryptographic key infrastructures.
For instance, the CAs, as the underlying public key infrastructure for SSL/TLS protocol, are responsible for issuing digital certificates that certify the ownership of a public key by the named principal of the certificate and allows others to rely upon signatures made by the private key corresponding to the certified public key.

Recent research have demonstrated that a variety of attacks \cite{basin2016design, yu2017decim} and mis-issuance problems \cite{kumar2018tracking, scheitle2018first} might cause complications during certificate issuance procedures.
For instance, Kumar et al. \cite{kumar2018tracking} analyze CAs' certificate mis-issuance incidents using a certificate linter named ZLint and Scheitle et al.  \cite{scheitle2018first} focus on issues with certification authority authorization (CAA) DNS records.
To further mitigate the threat of attack and mis-issuance,
notions of \textit{certificate transparency} \cite{laurie2014certificate,laurie2013certificate}, \textit{CertChain} \cite{chen2018certchain}, \textit{verifiable key directory} \cite{melara2015coniks,bonneau2016ethiks,chase2019seemless}, \textit{transparency overlay} \cite{chase2016transparency} and \textit{authority transparency} \cite{xu2020trustworthy} have been proposed. 
To be more precise, certificate transparency frameworks such as those presented in \cite{laurie2014certificate,laurie2013certificate} are intended to increase the transparency of users' certificates, whereas the CONIKS \cite{melara2015coniks,bonneau2016ethiks} and SEEMless \cite{chase2019seemless} are intended to be used with general key directories in end-to-end encryption systems.
Unlike the above-mentioned conventional transparency frameworks, which focus on the static binding of a public key and an identity, emerging authority transparency \cite{xu2020trustworthy} focuses on the dynamic key generation interactions in TPA that is requisite by modern cryptosystems such as attribute-based encryption (ABE) \cite{gorasia2016improving,agrawal2017fame}, functional encryption (FE) \cite{boneh2011functional,abdalla2018multi}, and multi-key homomorphic encryption (HE) \cite{ananth2020multi, chen2019multi}.

However, the initial and formal design of authority transparency has considerable limitations that hinder its deployment and application in several areas, such as emerging third-party service enabled privacy-preserving applications \cite{sans2018reading, xu2019cryptonn, xu2019hybridalpha, ryffel2019partially, xu2021fedv}.
Specifically, these limitations include: 
(\romannumeral1) the definitions and protocols designed in authority transparency model only work on, and relies on, the ABE cryptosystems; and
(\romannumeral2) the implementation of authority transparency framework is based on a secure logging system.
In short, existing authority transparency proposal does not directly support other emerging cryptosystems such as the FE and multi-key HE families that have been used to build secure computation protocols \cite{xu2019hybridalpha, xu2019cryptonn, ryffel2019partially, xu2020revisiting}.

Besides the \textit{identity-to-public-key-binding stealthy targeted attack} and the \textit{private-key-service censorship attack} as illustrated in \cite{xu2020trustworthy}, FE or multi-key HE based applications have additional privacy threats; for instance, there is a potential inference attack by manipulating a malicious functionality-related vector, as illustrated in \cite{xu2019hybridalpha,xu2019cryptonn}.
Furthermore, the deployment of secure logging system based authority transparency solution pose a challenge with regards to being broadly accepted by the Internet community because:
(\romannumeral1) it requires several commercial companies or non-profit organizations that have the computation and storage capabilities to deploy a publicly auditable secure logging system such as that used in the certificate transparency community (e.g., secure logging systems deployed by Google and Mozilla);
(\romannumeral2) there is also a lack of a concrete mechanism for the entities to participate in a transparency framework and monitor and audit \textit{unintentional} or \textit{malicious} behaviors.

To address the aforementioned limitations, in this paper, 
we propose an approach that provides transparency and trustworthiness of third-party authority and IaaS using blockchain techniques - in short, \texttt{TAB} - for emerging third-party service-enabled crypto-based privacy-preserving applications.
For simplicity, we use FE-based privacy-preserving systems proposed in \cite{xu2019hybridalpha, xu2019cryptonn, ryffel2019partially} as the underlying application to illustrate the \texttt{TAB} approach. 
In particular, to achieve the transparency and trustworthiness goal, \texttt{TAB} employs the Ethereum blockchain as the underlying public ledger infrastructure, and also includes a novel and well designed Ethereum smart contract to support automatic accountability with an additional incentive mechanism to motivate participants to participate in the auditing process and punish unintentional misbehaviors or malicious behaviors. 
We summarize our key \textit{\textbf{contributions}} as follows:

We first revisit the notion of \textit{authority transparency} model and propose our formal \texttt{TAB} model with new definitions and protocols to address the entity trust issues considering the scenarios of generic crypto-based privacy-preserving applications where the cryptographic infrastructure TPA or centralized key server is commonly assumed to be \textit{fully trusted} and the application entities (e.g., third-party IaaS and data sources as illustrated in \figurename~\ref{fig:application}) is usually assumed to be \textit{honest-but-curious}.
    
Next, we design a novel smart contract to achieve automatic accountability based on our design of the \texttt{TAB} model and employ the Ethereum blockchain as the underlying public ledger infrastructure. 
    
We also design an incentive mechanism in the smart contract to (\romannumeral1) reward a TPA if it fulfills its obligation; (\romannumeral2) punish any entity that violates its responsibility, and (\romannumeral3) encourage other entities to help audit and inspect the potential malicious behaviors caused by the \textit{assumed fully trusted} TPA and \textit{assumed honest} participants.
    
We finally analyze the security guarantee of \texttt{TAB} and present the experimental evaluation on the smart contract implemented in the Ethereum official test network - Rinkeby. The evaluation result shows that \texttt{TAB} is efficient and provides security and privacy guarantees.


\section{Background and Preliminaries}
\label{sec:bp}

Here, we briefly present preliminaries and background of related concepts such as functional encryption and its related applications, authority transparency, blockchain, Ethereum, and smart contract.

\subsection{TPA-based Cryptosystems and Applications}
\label{sec:bp:fe}

Emerging modern cryptographic schemes, especially those that rely on a \textit{third-party authority (TPA)} to provide key services, are being adopted in privacy-preserving applications, where data is encrypted and the data management operations such as querying, accessing control, and computation are over the encrypted data. 
A TPA is a critical component in these cryptosystems, and it is generally assumed to be fully trusted.
Such an assumption is very common in cryptography research community. However, deploying such a trusted TPA component in a real scenario is still a challenge because there is a lack of (\romannumeral1) incentive mechanisms to encourage a participant (i.e., a third-party entity) to play the role of the \textit{authority} and (\romannumeral2) a transparent mechanism to ensure that such a TPA works as expected when considering the attacks such as \textit{identity-to-public-key-binding stealthy targeted} attack and \textit{private-key-service censorship attack} as illustrated in \cite{xu2020trustworthy}.

Beyond \textit{authority transparency} that addresses the trust issues in a TPA caused by the aforementioned attacks, we address incentive issues related to participants' engagement in a transparency framework via Ethereum blockchain techniques, and tackle additional privacy leakage issue caused by assuming that participants are \textit{hones-but-curious} in crypto-based privacy-preserving applications \cite{xu2019cryptonn,xu2019hybridalpha}.
Our proposed blockchain-based \texttt{TAB} framework can support various TPA-based cryptosystems such as ABE-enabled applications as illustrated in \cite{xu2020trustworthy} and the emerging FE-enabled privacy-preserving applications.
Here, we briefly introduce the FE cryptosystem and FE-based applications; we also differentiate between the ABE and FE structures/components.

\subsubsection{Functional Encryption (FE)}
FE is a generalization of public-key encryption in which any party with an issued functional secret key allows us to compute a function of what a ciphertext is encrypting.
A FE scheme for functionality $\mathcal{F}$ is a tuple $\mathcal{E}_{FE}$ = \textit{(Setup, KeyDerive, Encrypt, Decrypt)} of four algorithms \cite{boneh2011functional,abdalla2018multi}, where the \textit{Setup} and \textit{KeyDerive} algorithms are run by a TPA that is assumed to be \textit{fully trusted}.
A data owner can adopt the \textit{Encrypt} algorithm to protect its data, while a data user with the functional decryption key issued by its TPA can compute the function over the ciphertext to acquire the function result without learning the original data via \textit{Decrypt} algorithm.







\subsubsection{FE-based Application and Potential Privacy Leakage}
The feature of computing over encrypted data makes functional encryption a promising approach for employing secure multi-party protocols for privacy-preserving machine learning (PPML) \cite{xu2019hybridalpha,xu2019cryptonn}.
While employing FE, a PPML also inherits the assumption of a \textit{trusted} TPA.
Besides, PPML techniques typically assume that the aggregator or coordinator (that is, the decryption party when PPML uses a FE scheme) is \textit{honest-but-curious}. 

Security guarantee provided by a FE scheme can ensure that the encrypted data cannot be compromised by an adversary \cite{abdalla2018multi}.
However, there is still potential privacy leakage in PPML approaches that use FE schemes, as demonstrated in \cite{xu2019hybridalpha,xu2019cryptonn}; here, an authorized \textit{honest-but-curious} decryption party may exploit a manipulated vector to request a functional decryption key to repeatedly execute the decryption algorithm over the encrypted data and store the intermediate data to infer partial information in the encrypted data.
For the specific inference attack, we refer the readers to \cite{xu2019hybridalpha,xu2019cryptonn} for more details.

\subsubsection{Comparing ABE and FE}
\label{sec:bp:diff}
ABE is also a type of public-key encryption in which ciphertexts are dependent upon access policy over a set of attribute credentials (e.g., age, affiliation, etc.) and any party with proper attribute credentials can be issued a secret key to access the encrypted data.
A \textit{(ciphertext-policy) attribute-based encryption (ABE)} scheme for access policy $\mathcal{A}$ is a tuple $\mathcal{E}_{ABE}$ = \textit{(Setup, KeyGeneration, Encrypt, Decrypt)} of four algorithms \cite{goyal2006attribute, bethencourt2007ciphertext}.
As in ABE, the \textit{Setup} and \textit{KeyGeneration} algorithms are run by a TPA that is assumed to be \textit{fully trusted}. A data owner uses the \textit{Encrypt} algorithm with a specified access policy to protect her data, while the data user with proper attribute credentials that satisfy the access policy can access (Decrypt) the encrypted data.







The main difference between ABE and FE is the credentials that are used to generate or derive the private key. In ABE, the private key is generated based on a set of attributes of a data user, while the functional decryption key is derived from a function-related vector in FE for the functionality of inner-product scheme.
Besides, the adoption of FE may introduce potential inference threats as illustrated in \cite{xu2019hybridalpha}.

\noindent\textit{\textbf{Remark}}. Unlike  \textit{authority transparency} \cite{xu2020trustworthy} that builds on the ABE scheme, for simplicity, in this paper, we use the recently proposed FE-based applications \cite{xu2019hybridalpha,xu2019cryptonn, ryffel2019partially} as underlying examples to illustrate the key features of \texttt{TAB}.
Specifically, \texttt{TAB} focuses on providing transparency in cases related to above-discussed assumptions, namely, a \textit{\textbf{trusted}} TPA and a \textit{\textbf{honest-but-curious}} participants, to increase users' trust in a system.
In Section~\ref{sec:tab:applicability}, we analyze the applicability of \texttt{TAB} in other TPA-based cryptosystems.

\subsection{Authority Transparency}
\label{sec:bp:at}
Authority transparency is defined as a publicly auditable set of a TPA's activities. 
The goal is to ensure that a TPA fulfills its auditing obligations ($\mathcal{O}$) related to public parameter distribution ($\mathcal{O}_{pp}$) and trustworthy key service ($\mathcal{O}_{ks}$), continuously and transparently.
We formally define authority transparency as below; here, we adopt the notation from \cite{bellare2015interactive}. 

\begin{definition}[Authority Transparency \cite{xu2020trustworthy}]
    Let $\mathcal{T}, \mathcal{L}$ and $\mathcal{C}$ denote a third-party authority, a log server, and a client, respectively, that use a set of interactive protocols.
    Let $\mathcal{C}.actor, \mathcal{C}.auditor$ and $\mathcal{C}.monitor$ represent the roles of the actor, auditor, and monitor that execute the application, auditing, and monitoring modules, respectively.
    We define \textit{authority transparency}, $\mathcal{AT}^{\mathcal{T},\mathcal{L},\mathcal{C}}_{\mathcal{O}}$,  as a set of six \textit{interactive} protocols:
    $$\mathcal{AT}^{\mathcal{T},\mathcal{L},\mathcal{C}}_{\mathcal{O}} = (\text{Gen}_{\mathcal{O}}, \text{Log}_{\mathcal{O}_{pp}}, \text{Log}_{\mathcal{O}_{ks}}, \text{Check}_{\mathcal{O}}, \text{Inspect}, \text{Gossip}), $$
    and each protocol is defined as follows:
    {\footnotesize
    \begin{align}
        &(S_{\mathcal{O}_{pp}}, S_{\mathcal{O}_{ks}}) \gets \text{Run}(\secparam, \text{Gen}_{\mathcal{O}}, \set{\mathcal{T}, \mathcal{C}.actor}, (\varepsilon, \varepsilon)) \\
        &(b_{\mathcal{T}}, \varepsilon) \gets \text{Run}(\secparam, \text{Log}_{\mathcal{O}_{pp}}, \set{\mathcal{T}, \mathcal{L}}, (S_{\mathcal{O}_{pp}}, \varepsilon)) \\
        &(b_{\mathcal{T}}, b_{\mathcal{C}}, \varepsilon) \gets \text{Run}(\secparam, \text{Log}_{\mathcal{O}_{ks}}, \set{\mathcal{T}, \mathcal{C}.actor, \mathcal{L}}, (\varepsilon, \mathcal{O}_{ks}.\mathcal{S}_{\mathcal{C}}, \mathcal{O}_{ks}.\mathcal{S}_{\mathcal{T}})) \\
        &(\varepsilon, b_{\mathcal{C}.auditor}) \leftarrow \text{Run}(\secparam, \text{Check}_{\mathcal{O}}, \set{\mathcal{L},\mathcal{C}.auditor}, (\varepsilon, \varepsilon)) \\
        &(b_{\mathcal{L}}, \varepsilon) \leftarrow \text{Run}(\secparam, \text{Inspect}, \set{\mathcal{L}, \mathcal{C}.monitor}, (\varepsilon, \varepsilon)) \\
        &(\text{evidence}) \leftarrow \text{Run}(\secparam, \text{Gossip}, \set{\mathcal{C}.auditor, \mathcal{C}.monitor}, (\varepsilon, \varepsilon))
    \end{align}
    }
\end{definition}

The order of parameters in the input tuple and the order of elements in the output are consistent with participating entities. 
For instance, in protocol $(b_{\mathcal{T}}, \varepsilon) \leftarrow \text{Run}(\secparam, \text{Log}_{\mathcal{O}_{pp}}, \set{\mathcal{T}, \mathcal{L}}, (S_{\mathcal{O}_{pp}}, \varepsilon))$, there exists two participants: $\mathcal{T}$ has the input $\mathcal{O}_{pp}$ while $\mathcal{L}$ has no input, as denoted by $\varepsilon$.

We briefly introduce each interactive protocol as follows: 
\begin{itemize}
    \item[(1)] $\textit{Gen}_{\mathcal{O}}$ is a protocol between $\mathcal{T}$ and $\mathcal{C}.actor$ that generates the audit obligations to be logged; 
    \item[(2)] $\textit{Log}_{\mathcal{O}_{pp}}$ is a protocol between $\mathcal{T}$ and $\mathcal{L}$ that is used to record $\mathcal{O}_{pp}$ in the public log; 
    \item[(3)] $\textit{Log}_{\mathcal{O}_{ks}}$ is a protocol involving $\mathcal{T}$, $\mathcal{L}$ and $\mathcal{C}.actor$ that is used to record $\mathcal{O}_{ks}$ in the public log; 
    \item[(4)] $\textit{Check}_{\mathcal{O}}$ is a protocol involving $\mathcal{L}$, $\mathcal{C}.actor$ and $\mathcal{C}.auditor$ that is used to check whether or not an audit obligation $\mathcal{O}_{pp}$ or $\mathcal{O}_{ks}$is in the log;
    \item[(5)] $\textit{Inspect}$ is a protocol between $\mathcal{L}$ and $\mathcal{C}.monitor$ that is used to allow the monitor to inspect the contents of the log and find suspicious audit obligations $\{\mathcal{O}_{i}\}$; 
    \item[(6)] $\textit{Gossip}$ is a protocol between $\mathcal{C}$\textit{.auditor} and $\mathcal{C}$\textit{.monitor} that is used to compare different versions of a log and detect any inconsistencies caused by misbehavior of a participant or on behalf of the log server.
\end{itemize}

Unlike the authority transparency approach proposed in \cite{xu2020trustworthy} that is built on the secure logging system, our proposed \texttt{TAB} relies on the Ethereum blockchain.
Thus, the above-mentioned protocols are not directly applicable in our blockchain-based \texttt{TAB} framework.
We will present our relevant definitions in Section~\ref{sec:tab:spec}.

\subsection{Blockchain, Ethereum and Smart Contract}
\label{sec:bp:sc}
A blockchain is a growing list of records (a.k.a, blocks) that are linked via cryptographic techniques, where each block contains a cryptographic hash of the previous block, a timestamp, and the transaction data.
In particular, the blockchain is a public distributed database of records, transactions, or digital events that have been executed and shared among various participants. In our proposed work we, employ a blockchain as the underlying public ledger infrastructure instead of the secure logging system adopted in \cite{xu2020trustworthy}.

Ethereum is an open-source and public blockchain-based distributed computing platform supporting smart contracts \cite{wood2014ethereum}.
Usually, there are two types of accounts in Ethereum, namely External Owned Accounts (EOAs) controlled by private keys associated with users and Contract Accounts assigned to smart contracts.
A smart contract in Ethereum refers to a piece of code, for instance, a Solidity\footnote{https://github.com/ethereum/solidity} program code that usually consists of multiple functions, few parameters and perhaps some modifiers. 
To deploy a smart contract, an ordinary user can compile the contract to generate the corresponding bytecodes and application binary interface (ABI), and then send a contract creation transaction to the Ethereum network with the bytecodes and ABI.
Upon receiving a transaction, the miners of the Ethereum network will include the bytecodes into the newly coming block being added.
Each successfully deployed contract account can be viewed as a small decentralized computation and storage unit that can execute specific functions defined in the contract and also store data allowed by the contract.
As a result, the transactions, messages, as well as the inputs of the functions are all recorded by the Ethereum blockchain, and, hence, the outputs of the functions are deterministic because the distributed miners can ensure that.
Note that it is not free to either deploy a smart contract or to call a function of existing smart contracts in Ethereum.
A user needs to pay Gas\footnote{https://github.com/ethereum/wiki/wiki/Whisper} that can be exchanged with Ether, the cryptocurrency used in Ethereum.

\section{\textit{TAB} Framework}
\label{sec:tab}

\subsection{Overview of \textit{TAB}}
\label{sec:tab:overview}


\subsubsection{Entities in \textit{TAB}}

\begin{figure*}[!t] 
  \centering 
  \includegraphics[width=0.7\textwidth]{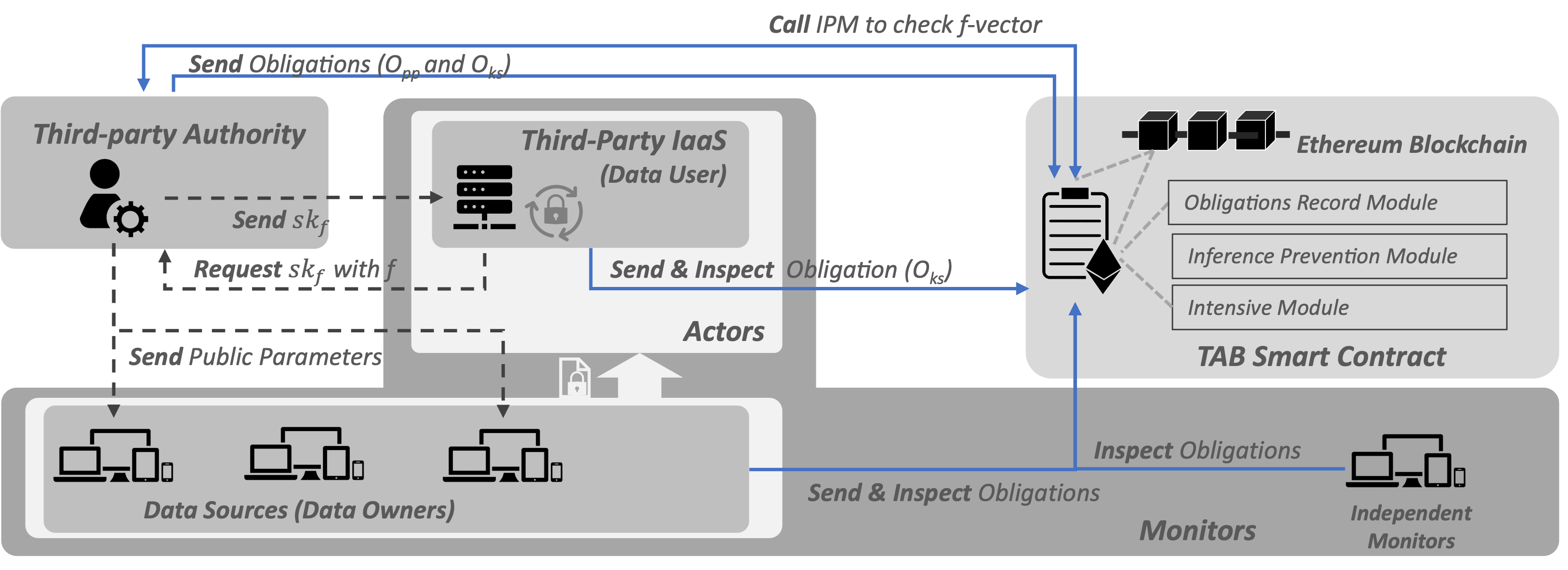}
  \caption{Overview of the \texttt{TAB} framework. Note that the dashed lines represent the procedures of the crypto-based privacy-preserving applications, while the solid lines denote the procedures of \texttt{TAB} framework.}
  \label{fig:tab:overview} 
\end{figure*}

\figurename~\ref{fig:tab:overview} illustrates the architecture of \texttt{TAB} framework. Note that the dashed lines represent the procedures of crypto-based privacy-preserving applications, while the solid lines denote the procedures of the \texttt{TAB} framework.
\texttt{TAB} consists of the following entities:

\noindent \textit{\textbf{TPA}}. The TPA is the same role as in the ordinary FE cryptosystem, but in \texttt{TAB} it has additional responsibilities to fulfill, including: (a) submitting the public parameters obligations (in particular, identity-to-public-key bindings), (b) reporting its fulfillment of obligations in the key service process, and (c) verifying that the submitted/reported obligations are permanently recorded in the blockchain.

\noindent \textit{\textbf{Actors}}. Actors include all users of an crypto-based privacy-preserving application, namely, the entities (e.g., \textit{data owner}) that employ the encryption algorithm and the entities (e.g., \textit{data user}) that perform secure computation or access control via the decryption algorithm. Besides, the actors may also need to fulfill the obligations of key service because they are involved in interaction with other actors and/or the TPA.

\noindent \textit{\textbf{Monitors}}. Monitors are responsible for inspecting the contents of the recorded auditing obligations to find suspicious obligations. In \texttt{TAB}, the encryption entities or the additional independent entities play the role of the monitors.

\noindent \textit{\textbf{Administrator}}. An administrator is responsible for the deployment, maintenance, and administration of the smart contract. 
The smart contract mainly includes three modules: (a) the obligation record module that provides various interaction functions for the entities to carry out recording, auditing and inspection requirements related to the obligations, (b) the incentive mechanism that provides the payment and reward functions to the participants, and (c) the \textit{inference prevention module (IPM)}, previously deployed in a TPA as illustrated in \cite{xu2019hybridalpha}.
Note that the Ethereum blockchain can ensure the trustworthiness of smart contracts; it can also ensure that the recorded obligations are distributed, open, and tamper-proof.
Note that once the smart contract deployed it does not need a centralized administration. 

\subsubsection{Notations and Use Scenarios}
To elaborate our \texttt{TAB}, we first present the notations, entities, and scenarios of applying our \texttt{TAB} framework in a crypto-based privacy-preserving environment.
Here we use FE as the underlying crypto scheme to present \texttt{TAB} framework. 
In Section~\ref{sec:tab:applicability}, we analyze the applicability of \texttt{TAB} in other TPA-based cryptosystems.
Suppose that we have a group of data owners $\{\mathcal{C}^{\textit{owner}}_{i}\}_{i\in[n]}$ that will share their private data $\pmb{x} = \{x_i\}_{i\in[n]}$ encrypted by an FE scheme where for simplicity we assume that $\mathcal{C}^{\textit{owner}}_{i}$ owns data $x_i$ , a group of data users $\{\mathcal{C}^{\textit{user}}_{j}\}_{j\in[m]}$, where each data user has a vector $\pmb{y}_{j}$ and needs to acquire the inner-product $\langle\pmb{x},\pmb{y}_{j}\rangle$ over the ciphertext of $\pmb{x}$, and a TPA $\mathcal{A}$ that provides public and private key services for these data owners and users.
Furthermore, let $\{\mathcal{C}^{\textit{owner}}_{k}\}_{k\in[l]}$ be the monitors.
We use $\mathcal{B}$ to represent the Ethereum blockchain, and let $\mathcal{B}^{\textsc{TAB}}_{\textsc{SC}}$ denotes our proposed smart contract deployed in the blockchain.

\subsection{Threat Model}
\label{sec:tab:threat}

\noindent\textbf{Threat Model of Privacy-Enhanced Applications}:
Existing crypto-based privacy-preserving applications are usually based on some common assumptions: (\romannumeral1) a centralized TPA or key server is assumed to be \textit{fully trusted} and (\romannumeral2) both decryption and encryption entities are assumed to be \textit{honest-but-curious}. 
Hence, the threat models in such cases typically focus on an adversary who attempts to compromise the encrypted data and  a\textit{curious} entity that launches potential privacy attacks (e.g., infer the private information), while \textit{honestly} following the protocols/algorithms.

\noindent\textbf{Threat Model of \texttt{TAB}}:
As illustrated by the \textit{identity-to-public-key-binding stealthy targeted attack} and the \textit{private-key-service censorship attack}  in \cite{xu2020trustworthy}, a TPA or key server may not be trusted because of its \textit{unintentional misbehaviors} and/or \textit{malicious behaviors}. 
Similarly, the \textit{honest-but-curious} entities may also behave dishonestly.
In addition to addressing above threats, \texttt{TAB} focuses on increasing entities' trust on the TPAs and other \textit{honest-but-curious} entities through transparency approach.
In particular, \texttt{TAB} mitigates the dependence of crypto-based privacy-preserving applications on the assumptions of a \textit{trusted} TPA or key server and an \textit{honest} entities.

To be more precise, we assume that such a dishonest adversary may pretend to behave honestly without being detected by other entities. Adversaries may not follow the specifications in the protocols, and/or attempt to conceal their activities.
In general, the \textit{dishonest} adversary includes the TPA and actors, where a dishonest TPA may attempt to forge a key service \textit{proof-of-work} without actually providing a valid key service; and, a dishonest actor may try to incorrectly blame other entities for misbehavior. 
Note that misbehavior may be related to non-malicious misuse by normal actors or the behavior of compromised actors controlled by an attacker.

We note that the case of potential collusion between a \textit{dishonest} TPA and \textit{honest-but-curious} actors is not fully considered in this paper. Rather than forbidding or preventing collusion through technical means, such collusion between two stakeholders (i.e., TPA and actors) can be solved by resorting to game theory and incentive mechanism designed in the smart contracts \cite{dong2017betrayal}.
\texttt{TAB} also involves the incentive mechanism from the aspects of each entity, and hence it can prevent such collusion in a game theory manner. 
We will not discuss that in the reset of the paper, and readers can refer to \cite{dong2017betrayal} for more details.

Furthermore, unlike the secure logging system based \textit{authority transparency} framework in \cite{xu2020trustworthy}, where the \textit{logger} is treated as a potential dishonest adversary, in \texttt{TAB}, the Ethereum smart contract is adopted as the public ledger infrastructure that has been proved to be a trusted computation platform.

\subsection{Proposed \textit{TAB} Framework}
\label{sec:tab:spec}

\subsubsection{\textit{TAB} Model}

Unlike the authority transparency approach in \cite{xu2020trustworthy} that builds on the secure logging system for ABE cryptosystem,
\texttt{TAB} uses the Ethereum blockchain, and to keep consistency, we adopt the similar concepts/notions of the authority transparency but it considers generic crypto-based privacy-preserving scenarios including emerging FE-based applications and a blockchain-based public ledger infrastructure.


Suppose that each entity $e$ in \texttt{TAB} is issued or self-generates an identity-based public and private key pair $\langle \text{pk}_e, \text{sk}_e \rangle$. 
Note that the key service interaction occurs between entity $\mathcal{C}^{\text{actor}}$ and authority $\mathcal{A}$, where each entity has already received its public and private key pair. For instance, let $\langle\text{pk}_{\text{actor}},\text{sk}_{\text{actor}} \rangle$ and $\langle \text{pk}_{\text{TPA}}, \text{sk}_{\text{TPA}} \rangle$ represent the public/private key pairs of the actor and the TPA, respectively.
Here, we first present the notion of \textit{public parameter audit obligation} and \textit{key service audit obligation}, and then present the formal definition of \texttt{TAB}.

\begin{definition}[Public Parameter Audit Obligation (PPAO)]
    A PPAO $\mathcal{O}_{pp}$ of $e$ is a map structure as follows:
    \begin{equation*}
        \mathcal{O}^{e}_{pp} := \text{H}(e_{id}) : \langle e_{id}, \text{pk}_{e}, \text{Sig}_{\text{sk}_{e}}( e_{id}, \text{pk}_{e})\rangle,
    \end{equation*}
    where $e_{id}$ represents the descriptive identifier of $e$, $\text{H}(\cdot)$ is a hash function, $\text{pk}_{e}$ denotes the public key binding of entity $e$, and $\text{Sig}_{\text{sk}_{e}}$ is the signature using $\text{sk}_{e}$.
\end{definition}

\begin{definition}[Key Service Audit Obligation (KSAO)]
\label{def:ksao}
    A KSAO $\mathcal{O}^{\mathcal{C}^{\text{actor}}, \mathcal{A}}_{ks}$ is a map structure consisting of a pair of key service snapshots
    \begin{equation*}
        \mathcal{O}^{\mathcal{C}^{\text{actor}}, \mathcal{A}}_{ks} := \text{H}(\mathcal{C}^{\text{actor}}_{id}, \mathcal{A}_{id}, r): \langle\mathcal{S}_{\text{req}},\mathcal{S}_{\text{resp}}\rangle,
    \end{equation*}
    where each snapshot is a 4-tuple as follows:
    \begin{align*}
      \mathcal{S}_{\text{req}} &:=  \text{H}(\mathcal{C}^{\text{actor}}_{id}, \mathcal{A}_{id}, r): \langle r, f, t_{\mathcal{C}^{\text{actor}}}, \text{Sig}_{\text{sk}_{\text{actor}}}(r, f, t_{\mathcal{C}^{\text{actor}}})\rangle, \\
      \mathcal{S}_{\text{resp}} &:= \text{H}(\mathcal{C}^{\text{actor}}_{id}, \mathcal{A}_{id}, r): \langle r, \sigma, t_{\mathcal{A}}, \text{Sig}_{\text{sk}_{\text{TPA}}}(r, \sigma, t_{\mathcal{A}})\rangle,
    \end{align*}
    such that 
    \begin{align*}
        \mathcal{S}_{\text{req}}.\text{H}(\mathcal{C}^{\text{actor}}_{id}, \mathcal{A}_{id}, r) &= \mathcal{S}_{\text{resp}}.\text{H}(\mathcal{C}^{\text{actor}}_{id}, \mathcal{A}_{id}, r) \\
        \mathcal{S}_{\text{resp}}.t_{\mathcal{A}}  - \mathcal{S}_{\text{req}}.t_{\mathcal{C}^{\text{actor}}} &> 0 \\
        \mathcal{S}_{\text{req}}.t_{\mathcal{A}} - \mathcal{S}_{\text{resp}}.t_{\mathcal{C}^{\text{actor}}} &< \delta_{t}
    \end{align*}
    where $r$ is a nonce selected by the key service requester, $t$ is the timestamp of key service processed by each entity, $f$ denotes the request content such as function related vector, $\sigma$ represents the proof-of-work that TPA has issued the key, $\delta_{t}$ is the threshold of timestamp difference indicating the
expected time of processing of the key service request by the TPA.
\end{definition}

\noindent\textit{\textbf{Remark}}.
In particular, $\mathcal{O}_{pp}$ is an \textit{identity-to-public-key} binding with the issuer's signature, while $\mathcal{O}^{\mathcal{C}^{\text{actor}}, \mathcal{A}}_{ks}$ is the \textit{proof-of-key-service}.
In the $\mathcal{O}^{\mathcal{C}^{\text{actor}}, \mathcal{A}}_{ks}$, for simplicity, to provide the \textit{proof-of-work} of issuing the functional decryption key $\text{sk}_{f}$ for the function related materials $f$, let $\text{Sig}ma$ be $\text{H}(\text{sk}_{f})$.

Based on the notion of \textit{public parameter audit obligation} and \textit{key service audit obligation}, we present the formal \texttt{TAB} model as follows:
\begin{definition}[\texttt{TAB} Model]
    Let $\mathcal{A}, \mathcal{B}$ and $\mathcal{C}$ denote a third-party \underline{A}uthority, a \underline{B}lockchain, and an \underline{A}ctor, respectively, which are parties involved in the interactive protocols.
    Let $\mathcal{C}.actor$ and $\mathcal{C}.monitor$ represent the roles of the actor and monitor that execute the functional and monitoring modules, respectively.
    We define \texttt{TAB} model, $\mathcal{M}$, as a set of five interactive protocols:
    \begin{equation*}
        \mathcal{M}^{\mathcal{A},\mathcal{B},\mathcal{C}}_{\mathcal{O}} = (\text{Gen}_{\mathcal{O}}, \text{Log}_{\mathcal{O}_{pp}}, \text{Log}_{\mathcal{O}_{ks}}, \text{Inspect}), 
    \end{equation*}
    and each protocol is defined as follows:
    {\footnotesize 
    \begin{align*}
        (S_{\mathcal{O}_{pp}}, S_{\mathcal{O}_{ks}}) &\gets \text{Run}(1^\lambda, \text{Gen}_{\mathcal{O}}, \{\mathcal{A}, \mathcal{C}.actor\}) \\
        (b_{\mathcal{A}}, \varepsilon) &\gets \text{Run}(1^\lambda, \text{Log}_{\mathcal{O}_{pp}}, \{\mathcal{A}, \mathcal{B}\}, (S_{\mathcal{O}_{pp}}, \varepsilon)) \\
        (b_{\mathcal{A}}, b_{\mathcal{C}}, \varepsilon) &\gets \text{Run}(1^\lambda, \text{Log}_{\mathcal{O}_{ks}}, \{\mathcal{A}, \mathcal{C}.actor, \mathcal{B}\}, (\mathcal{O}_{ks}.\mathcal{S}_{\mathcal{A}}, \mathcal{O}_{ks}.\mathcal{S}_{\mathcal{C}}, \varepsilon)) \\
        (b_{\mathcal{B}}, \varepsilon) &\gets \text{Run}(1^\lambda, \text{Inspect}, \{\mathcal{B}, \mathcal{C}.monitor\}, (\varepsilon, \varepsilon))
    \end{align*}
    }%
\end{definition}

Theorem~\ref{theorem:security} presents similar security guarantee as used in \cite{xu2020trustworthy}.
We present the details in Section~\ref{sec:eval:sp}.
\begin{theorem}
    \label{theorem:security}
    If the hash function is collision-resistant and the signature scheme is unforgeable, then \texttt{TAB} model comprises a secure transparency framework.
\end{theorem}

\noindent\textit{\textbf{Remark}}. Note that the formal definition of our \texttt{TAB} model is inherited from the \textit{authority transparency} model \cite{xu2020trustworthy} with needed changes considering the underlying Ethereum blockchain infrastructure.
Specifically, in the authority transparency model, the \textit{gossip} protocol essentially ensures the consistency of distributed logs without being tampered by an adversary, while the \textit{check} protocol guarantees that the submitted obligations are recorded by the logging system.
As \texttt{TAB} adopts the Ethereum blockchain as the underlying public ledger infrastructure, there is no need to run the \textit{gossip} and \textit{check} protocols because these logging-related functions are the features provided by the Ethereum smart contract.

\subsubsection{Design of \textit{$\mathcal{B}^{\textsc{TAB}}_{\textsc{SC}}$}}
\begin{figure}[!t] 
  \centering 
  \includegraphics[width=0.45\textwidth]{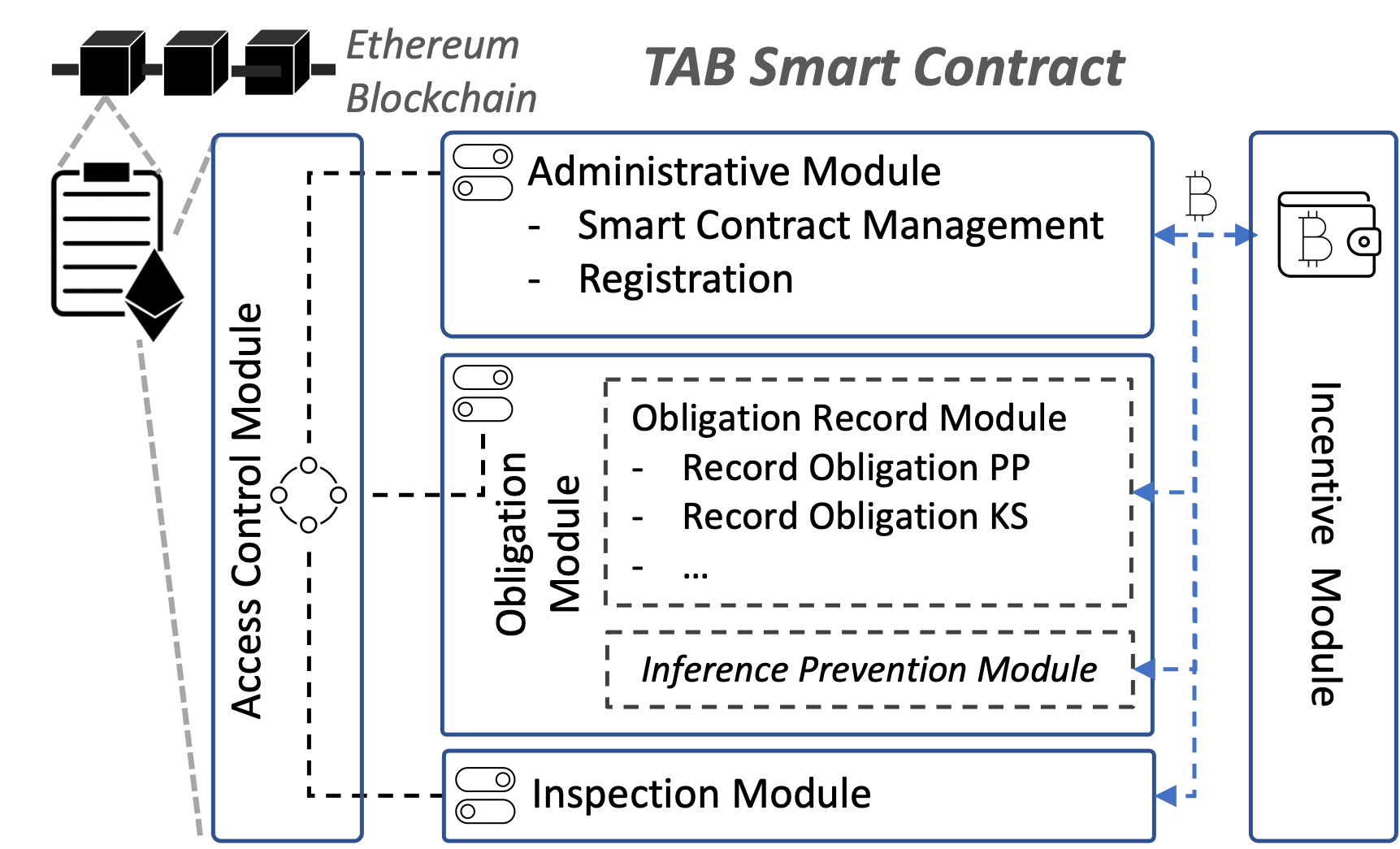}
  \caption{Overview of the \texttt{TAB} Smart Contract Interface.}
  \label{fig:tab:sc} 
\end{figure}

The \texttt{TAB} smart contract is a critical component in our framework. To support the goal of \texttt{TAB} framework, $\mathcal{B}^{\textsc{TAB}}_{\textsc{SC}}$ includes various types of modules: \textit{administrative module}, \textit{access control module}, \textit{obligation module}, \textit{inspection module}, and \textit{incentive module}.
As illustrated in \figurename~\ref{fig:tab:sc}, the access control module verifies a user's role within an application and determines whether or not to execute corresponding modules or functions.
The administrative module keeps track of the \texttt{TAB} smart contract's status and registration procedures.
The obligation module is concerned with the collection of various obligation records and the prevention of inference in privacy-preserving applications.
The inspection module enables authorized users to examine misbehavior or malicious activity.
The incentive module coordinates the above modules in order to reward or punish users using the Ethereum network's payment functionalities.
We discuss each module in details next.

\noindent\textbf{Administrative module}. This module allows the administrator role to deploy the smart contract into the Ethereum network. The module also includes functions such as opening and locking the enrollment, and allowing the participants to drop out.

\noindent\textbf{Access control module}. This module supports a basic role based access control (RBAC) mechanism that allows the account (a.k.a, the participating entities) have role-related permissions to call various functions.
In $\mathcal{B}^{\textsc{TAB}}_{\textsc{SC}}$, we define four types of roles: the \textit{TPA}, the \textit{actors of data owner}, the \textit{actors of data user}, the \textit{monitors} and the \textit{administrator} (i.e., the smart contract owner).
The administrative entity that deploys the smart contract becomes the \textit{smart contract owner}.
The ownership can be transferred to a new account if necessary.
Besides, it is also possible to relinquish this administrative privilege, which is a common pattern after an initial stage when there is a need for a decentralized administration.
After the deployment, each entity needs to register to its role by calling the corresponding function before it can use the ordinary features of the smart contract.

\noindent\textbf{Obligation module}. This module assists in recording the \textit{audit obligation} into the public ledger.
It also publishes its \textit{identity-to-public-key} binding to the Ethereum blockchain, as illustrated above.
Note that the identity of the entity is the unique public address (i.e., 42 hex string characters without case-sensitivity) of the blockchain account, which is derived from the entity's private key.
With regards to the key service audit obligation procedure, the key service requestor (i.e., data owner) can call the corresponding function (that includes role verification) with a randomly generated request identifier, the key-related request parameters, and the corresponding signature.
The function then automatically analyzes the request parameters via the \textit{inference prevention module (IPM)}.
Note that the IPM, previously deployed inside the centralized trusted TPA in vanilla FE-based applications, is now deployed in the smart contracts in a decentralized trust setting in the \texttt{TAB}.
Upon receiving the key service request with the request identifier, the TPA first checks the verification result of IPM.
If the request passes the verification, the TPA will issue the functional decryption key and then publish a response snapshot to fulfill the key service obligation.

\noindent\textbf{Inspection module}. This module mainly inspects the completeness of a pair of the key service snapshots to check whether the TPA has fulfilled its key service obligation or not. 
Besides, it also allows checking for the published \textit{identity-to-public-key} binding.
Besides  the inspection module that can prevent potential misbehaviors, we have introduced the RBAC mechanism to prevent partial misbehaviors and malicious behaviors as each entity only will be allowed to call corresponding functions with limited privilege.

\noindent\textbf{Incentive module}. This module, as part of $\mathcal{B}^{\textsc{TAB}}_{\textsc{SC}}$, includes several functions to enforce the incentive mechanism, as depicted in \figurename~\ref{fig:tab:sc}.
The incentive mechanism is based on payment features of the Ethereum network, where the token can be exchanged for real currency. 
As illustrated in \figurename~\ref{fig:tab:sc}, we design several functions as "publicly payable", which indicates that the smart contract is able to receive the transaction value (e.g., the Ether) when the function is successfully called and executed.

In general, $m$ data users need to pay equally for the cost of calling the registration function for themselves as well as for $n$ data owners and the TPA.
Each data user also needs to pay for the cost of calling the \textit{request obligation record} function and that of calling the \textit{response obligation record} function by the TPA.
Additionally, there exists a mechanism to punish the misbehaviors and malicious activities by a \textit{dishonest} TPA and data users.
To achieve that, the data owners and the TPA first need to register and pay the cost by themselves.
The data users make a deposit equally for all the entities' registration cost after the enrollment phase.
Then, the data owners and the TPA can call the \textit{disposable} reward function to withdraw the registration cost.
Besides, we make the TPA and data users make a \textit{guaranteed deposit} after the registration phase.
The monitors can register and pay the cost by themselves, and then calls the inspection function to check the suspicious behaviors.
If monitors find the malicious behaviors, they will acquire the reward from a fine to the corresponding entity (i.e., the guaranteed deposit of the entity).
Without the guaranteed deposit, the corresponding entity is not allowed to operate in/join the system.
We discuss the quantitative analysis of the cost of each entity in $\mathcal{B}^{\textsc{TAB}}_{\textsc{SC}}$ in Section~\ref{sec:eval:exp}.

\subsubsection{\textit{TAB} Procedures}
\label{sec:tab:procedure}
\begin{figure*}[!t] 
  \centering 
  \includegraphics[width=\textwidth]{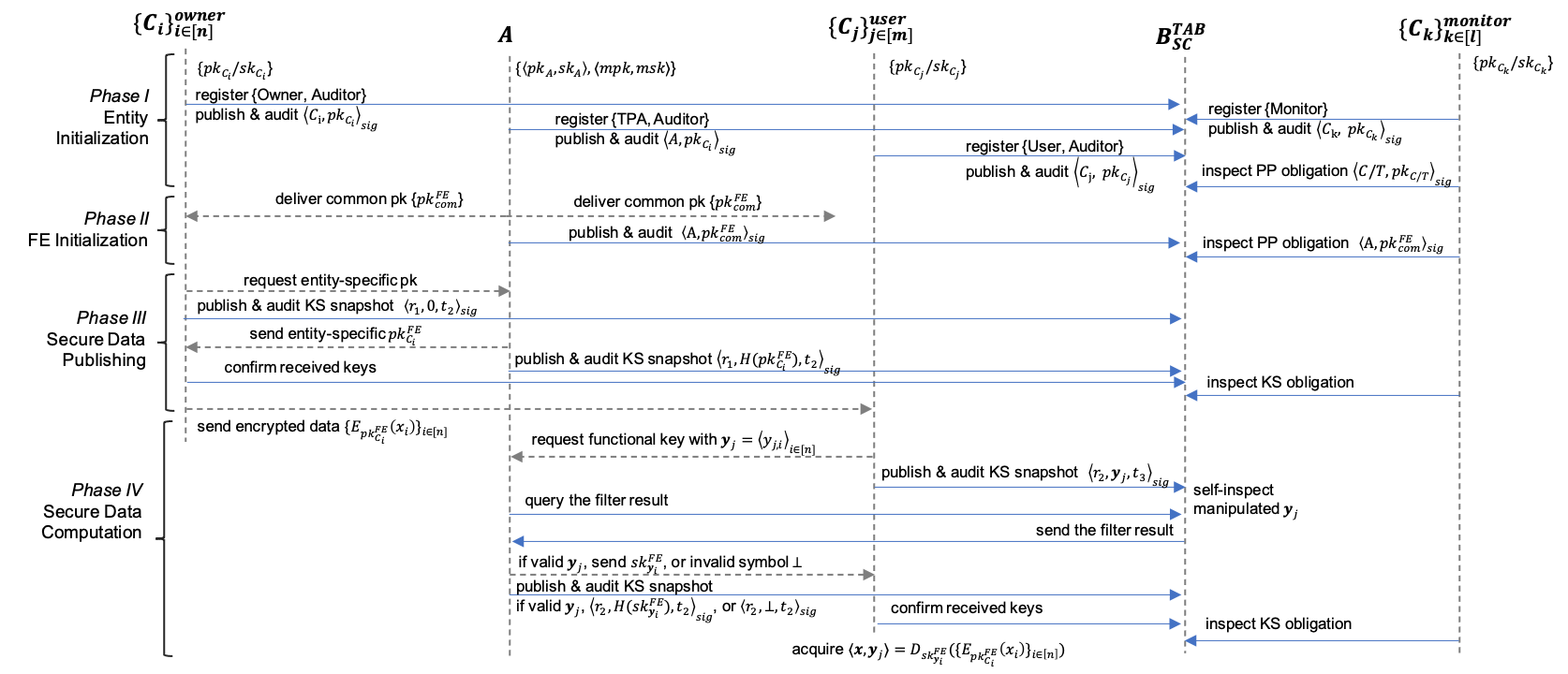}
  \caption{Illustration of the four phases with specific procedures in \texttt{TAB} in a FE-based privacy-preserving application scenario. 
  }
  \label{fig:tab:procedure} 
\end{figure*}

As depicted in \figurename~\ref{fig:tab:procedure}, we illustrate the four phases of the \texttt{TAB} framework with specific procedures in a typical FE-based privacy-preserving application scenario. 
Note that the dashed arrows represent the functional procedures of a typical FE-based application, while the solid arrows denote procedures specific to \texttt{TAB}.
In our design, each entity in the FE-based application can also play the role of the auditor and monitor, and we also allow additional monitors to help inspect the misbehaviors and malicious behaviors. Below, we present the specific procedures of each phase in  \texttt{TAB}.

\noindent\textit{Phase I: entity initialization}:
For each entity $e$ with role $e_{role}$ and identifier $e_{id}$ in the framework, it generates a public and private key pair $\langle\text{pk}_e,\text{sk}_e\rangle$.
Then, entity $e$ registers its role $e_{role}$ to $\mathcal{B}^{\textsc{TAB}}_{\textsc{SC}}$, and publishes its \textit{id-to-public-key binding } $\langle e_{id}, \text{pk}_{e}\rangle$ with its signature $\text{Sig}_{\text{sk}_{e}}(e_{id}, \text{pk}_{e})$ to $\mathcal{B}^{\textsc{TAB}}_{\textsc{SC}}$.

\noindent\textit{Phase II: FE (crypto) initialization}.
The TPA $\mathcal{A}$ sets up the FE cryptosystem with the master public key and master private key pair $\langle\text{mpk}^{\text{FE}},\text{msk}^{\text{FE}}\rangle$.
Using the master key, the TPA generates and sends the common public key $\text{pk}^{\text{FE}}_{com}$ for all entities (i.e., data owners and data users) in the FE-based application.
Then, the TPA publishes the binding $\langle\mathcal{A}_{id}, \text{pk}^{\text{FE}}_{com}\rangle$ with its signature $\text{Sig}_{\text{sk}_{\mathcal{A}}}(\mathcal{A}_{id},\text{pk}^{\text{FE}}_{com})$ to $\mathcal{B}^{\textsc{TAB}}_{\textsc{SC}}$.

\noindent\textit{Phase III: secure data publishing}.
For each data owner $\mathcal{C}^{\text{owner}}_i$, it first selects a nonce $r$ as the key service identifier. 
Then $\mathcal{C}^{\text{owner}}_i$ requests the entity-specific public key $\text{pk}^{\text{FE}}_{\mathcal{C}^{\text{owner}}_i}$ from the TPA with $r$.
Meanwhile, $\mathcal{C}^{\text{owner}}_i$ also sends a request key service snapshot $\mathcal{S}^{\mathcal{C}^{\text{owner}}_i}_{\text{req}}$ to $\mathcal{B}^{\textsc{TAB}}_{\textsc{SC}}$ as follows:
\begin{equation*}
    \mathcal{S}^{\mathcal{C}^{\text{owner}}_i}_{\text{req}} = \langle r, 0, t_{\mathcal{C}^{\text{owner}}_i}, \text{Sig}_{\text{sk}_{\mathcal{C}^{\text{owner}}_i}}(r, 0, t_{\mathcal{C}^{\text{owner}}_i})\rangle.
\end{equation*}
Then, the TPA generates $\text{pk}^{\text{FE}}_{\mathcal{C}^{\text{owner}}_i}$ for $\mathcal{C}^{\text{owner}}_i$ using its master keys, and also publishes a corresponding response key service snapshot $\mathcal{S}^{\mathcal{A}}_{resp}$ to $\mathcal{B}^{\textsc{TAB}}_{\textsc{SC}}$ to fulfill its key service audit obligation $\mathcal{O}^{\mathcal{C}^{\text{owner}}_i, \mathcal{A}}_{ks}$ with mapping key $\text{H}(\mathcal{C}^{\text{owner}}_{i,id}, \mathcal{A}_{id}, r)$ as follows:
\begin{equation*}
    \mathcal{S}^{\mathcal{A}}_{\text{resp}} = \langle r, \text{H}(\text{pk}^{\text{FE}}_{\mathcal{C}^{\text{owner}}_{i}}), t_\mathcal{A}, \text{Sig}_{\text{sk}_{\mathcal{A}}}(r, \text{H}(\text{pk}^{\text{FE}}_{\mathcal{C}^{\text{owner}}_{i}}),t_\mathcal{A})\rangle.
\end{equation*}
Each data owner then uses $\text{pk}^{\text{FE}}_{e^{\text{owner}}_i}$ to encrypt its data as follows:
$\{\text{Enc}_{\text{pk}^{\text{FE}}_{e^{\text{owner}}_i}}(x_i)\}_{i\in[n]}$.
Finally, the data owner publishes a receipt for the received $\text{pk}^{\text{FE}}_{e^{\text{owner}}_i}$.

\noindent\textit{Phase IV: secure data computation}.
Suppose that a data user $\mathcal{C}^{\text{user}}_{j}$ who has a vector $\pmb{y}_{j} = (y_1, ..., y_n)_{j}$ would apply inner-product functionality over the encrypted data $\{\text{Enc}(x_1), ..., \text{Enc}(x_n)\}$.
$\mathcal{C}^{\text{user}}_{j}$ also selects a key service identifier $r'$ first, and then requests the functional decryption key $\text{sk}^{\text{FE}}_{\pmb{y}_{j}}$ to the TPA with the vector $\pmb{y}_{j}$ and $r'$.
At the same time, $\mathcal{C}^{\text{user}}_{j}$ also sends the request key service snapshot $\mathcal{S}^{\mathcal{C}^{\text{user}}_{j}}_{\text{req}}$ to $\mathcal{B}^{\textsc{TAB}}_{\textsc{SC}}$ as follows:
\begin{equation*}
    \mathcal{S}^{\mathcal{C}^{\text{user}}_{j}}_{\text{req}} = \langle r', \pmb{y}_{j}, t_{\mathcal{C}^{\text{user}}_{j}}, \text{Sig}_{\text{sk}_{\mathcal{C}^{\text{user}}_{j}}}(r', \pmb{y}_{j}, t_{\mathcal{C}^{\text{user}}_{j}})\rangle.
\end{equation*}
Unlike the approaches proposed in \cite{xu2019cryptonn,xu2019hybridalpha} that deploy the \textit{inference prevention module (IPM)} within a TPA, we propose to deploy IPM in a smart contract as the TPA is not fully trusted in \texttt{TAB}.
Thus, the TPA needs to query $\mathcal{B}^{\textsc{TAB}}_{\textsc{SC}}$ to check the validity of $\pmb{y}_i$.
If $\pmb{y}_i$ is valid, the TPA generates $\text{sk}^{\text{FE}}_{\pmb{y}_{j}}$ for $\mathcal{C}^{\text{user}}_{j}$ using its master keys, and then publishes a corresponding response key service snapshot $\mathcal{S}^{\mathcal{A}}_{\text{resp}}$ to $\mathcal{B}^{\textsc{TAB}}_{\textsc{SC}}$ to fulfill its key service audit obligation $\mathcal{O}^{\mathcal{C}^{\text{user}}_{j}, \mathcal{A}}_{ks}$ with mapping key $\text{H}(\mathcal{C}^{\text{user}}_{j}, \mathcal{A}, r')$ as follows:
\begin{equation*}
    \mathcal{S}^{\mathcal{A}}_{\text{resp}} = \langle r', \text{H}(\text{sk}^{\text{FE}}_{\pmb{y}_{j}}), t_{\mathcal{A}}, \text{Sig}_{\text{sk}_{\mathcal{A}}}(r', \text{H}(\text{sk}^{\text{FE}}_{\pmb{y}_{j}}), t_{\mathcal{A}})\rangle.
\end{equation*}
Otherwise, the TPA refuses the key service and also publishes key service snapshot indicating that it has refused the key service, $\mathcal{S}^{\mathcal{A},\text{refuse}}_{\text{resp}}$, with refusing symbol $\perp$ to $\mathcal{B}^{\textsc{TAB}}_{\textsc{SC}}$ to fulfill its key service audit obligation as follows:
\begin{equation*}
    \mathcal{S}^{\mathcal{A},\text{refuse}}_{\text{resp}} = \langle r', \text{H}(\perp,\pmb{y}_{j}), t_{\mathcal{A}}, \text{Sig}_{\text{sk}_{\mathcal{A}}}(r', \text{H}(\text{sk}^{\text{FE}}_{\pmb{y}_{j}}), t_{\mathcal{A}})\rangle.
\end{equation*}
With the received $\text{sk}^{\text{FE}}_{\pmb{y}}$, a data user can compute the inner-product of $\langle\pmb{x},\pmb{y}\rangle$ by decryting as follows:
\begin{equation*}
    \langle\pmb{x},\pmb{y}_{j}\rangle = \text{Dec}_{\text{sk}^{\text{FE}}_{\pmb{y}_{j}}}(\{\text{Enc}_{\text{pk}^{\text{FE}}_{\mathcal{C}^{\text{owner}}_i}}(x_i)\}_{i\in[n]}).
\end{equation*}
Finally, the data owner publishes a receipt for the received $\text{sk}^{\text{FE}}_{\pmb{y}}$.

\noindent\textit{\textbf{Remark}}. To avoid redundant description, we do not present the roles of \textit{auditor} and \textit{monitor} in the above-mentioned procedures.
In particular, as illustrated in \figurename~\ref{fig:tab:procedure}, the \textit{data users}, \textit{data owners} and the \textit{TPA} also play the role of auditor that checks whether the audit obligations are recorded into the blochchain permanently. 
In our design, the \textit{data owners} also play the role of a monitor to check the suspicious obligations caused by misbehaviors and malicious behaviors from the TPA and adversarial \textit{data users}. 
For instance, as illustrated in \cite{xu2019cryptonn,xu2019hybridalpha}, an adversarial data user may infer the private vector $\pmb{x}$ by manipulating a vector to request the functional decryption key.
The monitor can inspect  $\mathcal{O}^{e^{\text{user}}, e^{\text{TPA}}}_{ks}$ to find the adversary's suspicious behaviors.
Furthermore, our design can also address the case of intentionally issuing an incorrect functional decryption key. For example, suppose the key request material (i.e., $\pmb{x}$) from the authorized user is correct after IPM checking, while the key is incorrect.
In \texttt{TAB}, the data user is also the monitor/auditor and can file a claim by manually calling the inspect function based on existing logged key service materials, where the inspection is based on the rule defined in the formal model presented above.

\subsection{Applicability of \textit{TAB}}
\label{sec:tab:applicability}

\texttt{TAB} is applicable to other popular TPA-based cryptosystems as well.
Specifically, we analyze the applicability of \texttt{TAB} in the attribute-based encryption (ABE) scheme that is the focus of \textit{authority transparency} proposed in \cite{xu2020trustworthy}.

Differences in key service audit obligations in FE and ABE schemes are as shown in \tablename~\ref{table:diff}.
The main difference in the key service is the credential type, namely, the function-related vector and the attributes that are usually represented in a character string.
These credentials are used to generate or derive the private key in the \textit{Setup} and \textit{KeyGeneration} phases.
As presented in Section~\ref{sec:tab:spec}, \texttt{TAB} is a general framework and is not restricted to the type of audit obligation that builds on different key service credentials.
Specifically, the request content $f$ and response proof-of-work $\sigma$ in Definition~\ref{def:ksao} are not limited to a function-related vector and corresponding hashed generated key as illustrated in Section~\ref{sec:tab:procedure}. 
\texttt{TAB} is applicable to ABE-based applications by replacing the following audit obligations:  
\begin{align*}
    \mathcal{S}^{\mathcal{C}^{\text{user}}_{j}}_{\text{req}} &= \langle r', \pmb{y}_{j}, t_{\mathcal{C}^{\text{user}}_{j}}, \text{Sig}_{\text{sk}_{\mathcal{C}^{\text{user}}_{j}}}(r', \pmb{y}_{j}, t_{\mathcal{C}^{\text{user}}_{j}})\rangle, \\
    \mathcal{S}^{\mathcal{A}}_{\text{resp}} &= \langle r', \text{H}(\text{sk}^{\text{FE}}_{\pmb{y}_{j}}), t_{\mathcal{A}}, \text{Sig}_{\text{sk}_{\mathcal{A}}}(r', \text{H}(\text{sk}^{\text{FE}}_{\pmb{y}_{j}}), t_{\mathcal{A}})\rangle,
\end{align*}
by the corresponding audit obligations:
\begin{align*}
    \mathcal{S}^{\mathcal{C}^{\text{user}}_{j}}_{\text{req}} &= \langle r', \pmb{S}_{j}, t_{\mathcal{C}^{\text{user}}_{j}}, \text{Sig}_{\text{sk}_{\mathcal{C}^{\text{user}}_{j}}}(r', \pmb{S}_{j}, t_{\mathcal{C}^{\text{user}}_{j}})\rangle, \\
    \mathcal{S}^{\mathcal{A}}_{\text{resp}} &= \langle r', \text{H}(\text{sk}^{\text{ABE}}_{\pmb{S}_{j}}), t_{\mathcal{A}}, \text{Sig}_{\text{sk}_{\mathcal{A}}}(r', \text{H}(\text{sk}^{\text{FE}}_{\pmb{S}_{j}}), t_{\mathcal{A}})\rangle,
\end{align*}
where $\pmb{S}_{j}$ is the attribute set and $\text{H}(\text{sk}^{\text{ABE}}_{\pmb{S}_{j}})$ is the corresponding access control private key generated by the TPA using the attribute set $\pmb{S}_{j}$ in the hash format.

\begin{table}[]
    \centering
    \begin{threeparttable}
    \caption{Different key service audit obligations in FE and ABE}
    \label{table:diff}
    \footnotesize
    \begin{tabular}{lll}
        \toprule
        Types & obligations in FE & obligations in ABE \\
        \midrule
        Setup-Output & private key & public parameter\\
        KeyGen-Input & function-related vector & attribute set\\
        KeyGen-Output & functional decryption key & attribute private key\\
        \bottomrule
    \end{tabular} 
    \end{threeparttable}
\end{table}

\section{Evaluation}
\label{sec:eval}

\subsection{Security, Privacy and Trustworthiness}
\label{sec:eval:sp}

\subsubsection{Security Guarantee} 
The security for the transparency framework is defined in terms of three properties \cite{xu2020trustworthy, chase2016transparency}: 
(\romannumeral1) \textit{log-consistency} - a dishonest public ledger cannot remain undetected if it tries to present inconsistent versions of the recorded obligations;
(\romannumeral2) \textit{unforgeable-service} - a dishonest TPA cannot forge a key service by sending valid key service snapshots, but not provide the key service to the actors;
(\romannumeral3) \textit{non-fabrication} - a dishonest TPA or actors cannot blame the public ledger for misbehavior if it has behaved honestly, and dishonest actors cannot prove the TPA for misbehavior if it has behaved honestly.

We note that \textit{log-consistency} relies on the security properties of the Ethereum blockchain. 
The \textit{unforgeable-service} and \textit{non-fabrication} properties depend on the designed smart contract functions and the adopted signature scheme.
Here, we use the game simulation-based reduction methodology to prove Theorem~\ref{theorem:security}. 
\begin{proof}
 \texttt{TAB} is built on three fundamental security components: the Ethereum blockchain as the public ledger infrastructure, the Secure Hash Algorithm 3 (SHA3) as the collision-resistance hash function, the Elliptic Curve Digital Signature Algorithm (ECDSA) to sign and validate the origin and integrity of messages. 
The security of three components has been proved in corresponding related work \cite{wood2014ethereum, dworkin2015sha, johnson2001elliptic}.
We only prove the above-mentioned three security properties.

\noindent\textit{Log-consistency}.
Unlike the existing transparency framework, \cite{xu2020trustworthy,chase2016transparency}, that relies on the customized public ledger, \texttt{TAB} uses the public blockchain that has already been proved to provide secure consistency feature \cite{wood2014ethereum}, and hence we do not present it here to avoid redundancy.

\noindent\textit{Unforgeable-service}.
In \texttt{TAB}, there are two possible issues related to forgeable-service: 
\begin{itemize}
    \item a dishonest TPA may publish $\mathcal{S}^{\mathcal{A}}_{\text{resp}}$ to the blockchain, but does not send the key $\text{sk}_{f}$ to the actors;
    \item the dishonest TPA may send an invalid key $\text{sk}^{'}_{f}$ to the actors, but publishes correct $\mathcal{S}^{\mathcal{A}}_{\text{resp}}$ generated from the valid key $\text{sk}_{f}$.
\end{itemize}
For the first issue, the confirmation phase of key service audit obligation cannot be accomplished in our designed smart contract, and then such adversarial behavior is easily detected by the monitors.
For the second issue, suppose that the dishonest TPA has the non-negligible advantage $\epsilon$ to break the \textit{unforgeable-service} security guarantee, and hence it can forge the hashed key component $\text{H}^{\text{SHA3}}(\text{sk}^{'}_{f})$ for $\text{sk}_{f}$ with advantage $\textbf{Adv}^{\mathcal{A}}_{\text{H}^{\text{SHA3}}(\text{sk}^{'}_{f}) \to \text{sk}_{f}} \ge \epsilon$.

To achieve that, the dishonest TPA hence needs the ability to find potential collision $\text{H}^{\text{SHA3}}(\text{sk}_{f}) = \text{H}^{\text{SHA3}}(\text{sk}^{'}_{f})$.
According to the security promise of SHA3, it is impossible to find that collision with non-negligible advantage \cite{dworkin2015sha}.
Thus, dishonest TPA does not have a non-negligible advantage to provide an unforgeable key service without being detected.
    
\noindent\textit{Non-fabrication}.
In \texttt{TAB}, a possible fabrication case is that dishonest actors may attempt to blame the TPA by publishing $\mathcal{S}^{\mathcal{C}^{\text{actor}}}_{req}$ to the blockchain but does not actually send the key request to the TPA.
Suppose that a dishonest actor has the non-negligible advantage $\epsilon$ to break the \textit{non-fabrication} security promise.
To launch the fabrication case, the dishonest actor needs to forge a fake $\mathcal{S}^{\mathcal{A}}_{\text{resp}}$ so that it can accomplish the confirmation phase. 
Thus, the dishonest actor is able to forge a fake signature of the TPA with advantage $\textbf{Adv}^{\mathcal{C}^{\text{actor}}}_{\text{sk}_{\mathcal{A}}} \ge \epsilon$.
However, it is impossible to break the ECDSA \cite{johnson2001elliptic}, as has been proved, namely, the unforgeability of the signature scheme.
Thus, the actors do not have a non-negligible advantage to frame up the TPA.
\end{proof}

\subsubsection{Privacy Guarantee}
Typically, authorized users request public and private keys from a TPA using (attribute) identities in privacy-preserving applications built on cryptosystems like as ABE, FE, and multi-key HE.
The major objective of \textit{authority transparency} \cite{xu2020trustworthy} and \texttt{TAB} is to audit the interactive key service discussed previously without invading the original key service.
As a result, privacy assurance of \texttt{TAB} focuses on preventing data leaking from audit materials.

Unlike the initial \textit{authority transparency} that focuses on ABE-based applications where partial attribute identities are privacy-sensitive, \texttt{TAB} also supports the privacy-enhanced computing applications that built on FE or multi-key HE schemes.
There is no privacy concern regarding the identity in an FE or multi-key HE because those identities could be any unique characters without any privacy-sensitive information.
For example, the identity of each entity in TAB could be the Ethereum network's public account address, which is a random 64-character hex string produced from the entity's private key.
Additionally, the TAB framework only receives the hash of attribute identifiers in ABE-based applications, not potentially sensitive attribute information.
As a result, such account identifiers or hash of auditing materials do not reveal any private identifiable information.

\subsubsection{Trustworthiness Goal}
The purpose of the  \texttt{TAB} approach is to deal with the trust issues raised by potential \textit{dishonest} entities by providing transparency.
\texttt{TAB} is able to prevent the attacks such as stealthy targeted attack and censorship attack as illustrated in \cite{xu2020trustworthy}.
Specifically, each dishonest entity needs to publish the key service snapshot to prove that it has fulfilled its obligation of public parameter distribution and private key service.
The designed smart contract can ensure that each entity's submitted audit obligations can be automatically cross-validated based on our designed protocols before being honestly and permanently recorded into the blockchain.
Our security analysis has shown that the misbehaviors or malicious behaviors of a TPA and the actors are easily detected.
Furthermore, the IPM is a critical component in FE-based applications \cite{xu2019cryptonn,xu2019hybridalpha} that helps to mitigate the inference threats.
In \texttt{TAB}, the IPM, which was deployed in a TPA in the scheme proposed in \cite{xu2019hybridalpha}, is moved to the smart contract, is automatically executed in a publicly auditable environment, and hence increase the transparency and trustworthiness of IPM.

\subsection{Experimental Evaluation}
\label{sec:eval:exp}

\subsubsection{Implementation and Setup}
The \texttt{TAB} model does not rely on the specific privacy-enhanced applications that are built on FE or multi-key HE cryptosystems, and hence for generality, we only present the evaluation on a pure \texttt{TAB} model with the simulated audit obligations where the key-related components are generated by the FE-based application in an off-line manner.

\noindent\textit{\textbf{Implemented Smart Contract}}:
We implement the smart contracts in \textit{Solidity} programming language using a \textit{Truffle} development environment, testing framework and asset pipeline for blockchains using the Ethereum Virtual Machine (EVM).
\figurename~\ref{fig:tab:sc} illustrates the core modules and interfaces of \texttt{TAB} smart contract, while the whole implementation is publicly available on Github \footnote{https://github.com/iRxyzzz/tab}.
We direct the reader to the Github repository for implementation specifics and discuss the fundamental implementation considerations below.
\texttt{TAB} mainly includes four types of functions as follows:

\noindent \textit{Access Control Modifiers}.
    The \textit{modifier} can be used to change the behavior of functions in a declarative way. In our implementation, we use the modifier to automatically check the privilege of each account that is defined in RBAC module prior to executing the function. We employ \textit{Ownable} and \textit{AccessControl} smart contracts from \textit{OpenZeppelin}\footnote{https://openzeppelin.com/contracts/} as the basis for our access control mechanism. To be specific, we define various access control modifiers in which the basic RBAC functions are integrated to satisfy our access control requirement.
    Except for the registration related functions, other functions are restricted by these modifiers.
    
\noindent \textit{Administrative and Incentive Functions}. 
    We define several administrative functions such as \textit{`enrollLock()'}, \textit{`enrollOpen()'}, \textit{`dropout()'} that allow the administrator to control the enrollment status. In \texttt{TAB}, each entity can register if and only if the enrollment is set as open by the administrator.
    After the enrollment is locked, the deposit operations are opened to the related entities.
    Besides, \texttt{TAB} also inherits the administrative functions such as \textit{`transferOwnership(newOwner)'}, \textit{`renounceOwnership()'}.
    These two functions allow transferring the ownership of the contract and leave the contract without owner, respectively.
    Furthermore, we also define several withdraw and deposit functions that help to establish a basis for the incentive and penalty mechanisms.
    
\noindent \textit{Registration Functions}.
    The registration functions mainly focus on the initialization phases of \texttt{TAB} (i.e., Phases I and II, as illustrated in Section~\ref{sec:tab:procedure}), where each entity is allowed to register a role, and publish its \textit{identity-to-public-key} binding in the blockchain.
    
\noindent \textit{Obligation Functions}.
    The obligation functions address the core features of the \texttt{TAB} model.
    As illustrated in Section~\ref{sec:tab:procedure} Phases III and IV, we use a three-phase commitment approach to achieve the obligation features.
    To be specific, \textit{`recordKSPKReq', and `recordKSSKReq'} allow the actors to publish the key service request snapshots, while \textit{`recordKSPKResp', and `recordKSSKResp'} allow the TPA to record corresponding key service response snapshots.
    Then, \textit{`recordKSPKResp'} function allows us to confirm the receipt of the key service.
    
\noindent \textit{Inspection Functions}.
    The inspection functions address the monitoring task for the recorded audit obligation as discussed in Section~\ref{sec:tab:procedure}.
    To be specific, \textit{`inspectObligationKS'} allows to automatically inspect the completeness of the key service obligations, while \textit{`inspectObligationPP'} permits the monitor to verify the published \textit{identity-to-public-key} binding.
    Regarding the incentive design, if a dishonest behavior is detected, the corresponding entity will be fined a fixed number of \textit{ethers} that will be provided as an incentive reward to the monitor.

\begin{table*}[]
    \centering
    \begin{threeparttable}
    \caption{The gas cost and test time of selected functions in various test case scenarios in the \texttt{TAB}.}
    \label{table:gas_cost}
    \footnotesize
    \begin{tabular}{clcrl}
        \toprule
        Test Cases & Functions & Gas Cost & Time & Description \\
        \midrule
        \multirow{4}{*}{Administrative} 
        & deployment & 4125603 & 183ms & deploy the smart contract  \\
        & enrollOpen & 44126 & 42ms & open the enrollment  \\
        & enrollLock & 14531 & 46ms & lock the enrollment \\
        & dropout & 28293 & 178ms & allow to drop out and withdraw the balance \\
        \hline
        \multirow{3}{*}{Incentive} 
        & depositGuarantee & 28083 & 48ms & deposit the guarantee  \\
        & rewardRegisterCost & 52949 & 43ms & reward registration cost for non-payable entity\\
        & rewardDeploymentCost & 51584 & 41ms & reward deployment for the administrator\\
        \hline
        \multirow{4}{*}{Registeration} 
        & registerAuthority & 38276 & 80ms & register the role of third-party authority \\
        & registerActorDataOwner & 38335 & 71ms & register the role of data owner \\
        & registerActorDataUser & 36555 & 70ms & register the role of data user \\
        & registerMonitor & 36521 & 72ms & register the role of monitor \\
        \hline
        \multirow{3}{*}{Obligation} 
        & recordKSSKReq & 43173 & 96ms & publish the key service request snapshot  \\
        & recordKSSKResp & 84211 & 55ms & publish the key service response snapshot \\
        & recordKSConfirm & 43402 & 49ms & confirm receipt of the key service obligation \\
        \hline
        \multirow{2}{*}{Inspection} 
        & inspectObligationKS & 24511 & 41ms & inspect the key service audit obligation \\
        & inspectObligationPP & 37482 & 46ms & check the correct of the public parameter \\
        \bottomrule
    \end{tabular} 
    \end{threeparttable}
\end{table*}

\noindent\textit{\textbf{Experimental Setup}}:
Our experiments have been performed on a Macbook Pro platform with 2.3GHz 8-Core Intel Core i9 processors and 32GB DDR4 memory.
Besides, we use the Ethereum official test network, Rinkeby, as the experimental environment to deploy our smart contract.
Furthermore, we write several \textit{JavaScript} test-cases using the automated testing framework of \textit{Truffle} that is built on Mocha\footnote{https://mochajs.org/} and provides a cleanroom environment.

Specifically, for demonstration, we use five Ethereum accounts to simulate various entities in \texttt{TAB}, namely, the role of the \textit{administrator}, the \textit{TPA}, the \textit{data owner}, the \textit{data user} and the \textit{monitor}.
With regards to various scenarios, we write corresponding test-cases to evaluate the performance (i.e., the gas cost and the time cost for scenarios such as administrative, registration, obligation, etc).

\subsubsection{Experimental Results}
We report the performance of \texttt{TAB} for selected functions for various test scenarios in \tablename~\ref{table:gas_cost}.
In particular, the performance includes two aspects: the gas cost and the test time. Gas is spent in Ethereum for deploying smart contracts or calling functions.
As reported in \tablename~\ref{table:gas_cost}, most functions cost very
little. 
Specifically, except for the smart contract deployment, the cost of each function is at the level of $10^5$ gas in general.
Regarding the most called functions for obligation and inspection, to record an audit obligation for one key service, the functions related to three-phase commitment (i.e., \textit{recordKSSKReq}, \textit{recordKSSKResp}, \textit{recordKSConfirm}) cost $3.7\times10^5$ gas, $8.4\times10^5$ gas, $4.3\times10^5$ gas, respectively. 
Besides, the cost of inspection for key service and public parameter audit obligations is $2.4\times10^5$ gas and $3.7\times10^5$ gas, respectively.

Furthermore, we also measure the time it takes to test the selected functions.
Except for the administrative functions, the calling time of rest of the functions is less than $100ms$.
Note that the time to test each function is measured in the  Ethereum test network.
The testing time is related to execution time instead of time taken to confirm the transaction .
Thus, the deployment time of the smart contract is only $183ms$ rather than the general time taken to confirm a transaction, namely, about 6 minutes.

\begin{figure}
    \centering
    \includegraphics[width=0.4\textwidth,trim=20 5 20 20, clip]{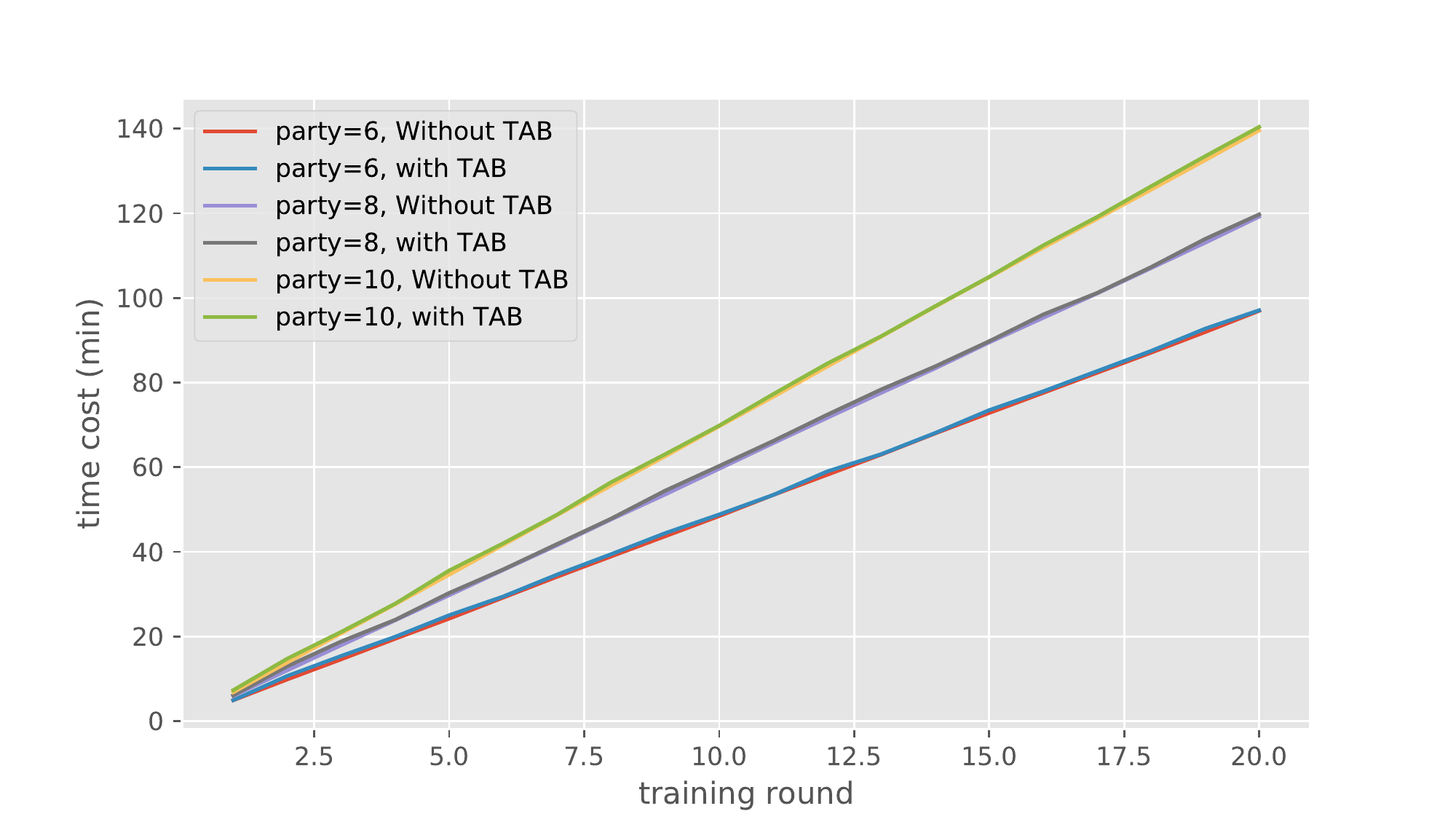}
    \caption{The time cost of \texttt{TAB}-enhanced privacy-preserving FL}
    \label{fig:scale}
\end{figure}

To further assess the \texttt{TAB} framework's scalability, we integrate it into a specific privacy-preserving federated learning application built on the FE cryptosystem with a TPA for key service, where the number of enrolled data owners (i.e., called party in the FL system) increased from 6 to 10 and the task of FL is to train a CNN model over MNIST dataset.
Due to the fact that TAB focuses exclusively on the key interactions between the user and the TPA, and the time cost is negligible in comparison to the time cost of FL training, the introduction of the \texttt{TAB} framework has slight effect on the time performance of the original privacy-preserving FL training, as shown in \figurename~\ref{fig:scale}.
Additionally, an increase in the number of participants has an effect on the amount of time spent on the supported privacy-preserving application (i.e., FL training), but not obvious on the amount of time spent on \texttt{TAB}.

\section{Related Work}
\label{sec:related}

\noindent\textbf{Privacy Enhanced Applications}.
Emerging FE schemes \cite{boneh2011functional, abdalla2018multi} have been shown to be a promising candidate for secure computation in privacy-preserving application scenarios where data is encrypted and outsourced, and the computation is carried out over the encrypted data.
Especially, recently proposed functional encryption for the functionality of computing the inner-product such as in \cite{abdalla2015simple,abdalla2018multi} raises the possibility of applying functional encryption in complex applications such as the federated learning and deep neural networks, as demonstrated in \cite{xu2019cryptonn,xu2021nn,xu2019hybridalpha,pokhrel2020federated,xu2021fedv,xu2020revisiting}.

Both HE and FE schemes are required to allow privacy-enhanced computing in crypto-based federated learning applications. While HE makes use of a centralized crypto dealer to synchronize key pairs, FE makes use of a third-party authority (TPA) to assist in the generation of public parameters and the provision of functional decryption keys for each function.
The crypto dealer or TPA is a vital component of those applications and is typically regarded to be \textit{fully trustworthy}.
Furthermore, the entities such as a coordinator in federated learning are also assumed to be \textit{honest-but-curious}.
However, the trust issues caused by malicious insiders in the TPA infrastructure \cite{xu2020trustworthy, chase2016transparency} and the privacy inference issues caused by curious entities \cite{xu2019cryptonn, xu2019hybridalpha} have not been investigated adequately.

\noindent\textbf{Transparency and Blockchain}.
The concept of transparency issues have received more and more attention due to malicious activities or misbehavior in various secure computing infrastructures and components.
For instance, the \textit{certificate transparency} proposed in \cite{laurie2013certificate, laurie2014certificate} aims to mitigate the certificate-based threats caused by fake or forged SSL certificates that are mistakenly or maliciously issued by insiders.
Most recent and related work such as \textit{CONIKS} \cite{melara2015coniks} and its following up work \textit{EthIKS} \cite{bonneau2016ethiks},  \textit{SEEMless} \cite{chase2019seemless}, and \textit{transparency overlay} \cite{chase2016transparency} target the key transparency in end-to-end encrypted communications systems and provide a formal transparency model.
The work closest to this proposed work is the \textit{authority transparency} framework proposed in \cite{xu2020trustworthy} that addresses the issues related to potentially dishonest TPAs in ABE-based applications using a secure logging based approach.
Further, blockchain based techniques have also been introduced to help increase the transparency of existing certificate transparency framework, such as in \cite{wang2020blockchain,chen2018certchain}.
However, there is still a lack of a mechanism to address the \textit{authority transparency} issues for emerging crypto-based privacy-preserving applications without relying on the complex secure logging systems. Such a transparency approach is important to ensure the trustworthy deployment of generic crypto-enabled systems.

\section{Conclusion}
\label{sec:conclusion}
This paper proposed the \texttt{TAB} framework to address transparency and trustworthiness of \textit{third-party authorities (TPAs)} and \textit{honest-but-curious} entities for generic modern crypto enabled privacy-preserving applications, as well as other schemes that have components similar TPAs and many entities that interact with them.
\texttt{TAB} employs the Ethereum blockchain as the underlying public ledger infrastructure and also incorporates a novel smart contract to support accountability with an additional incentive mechanism that motivates participants to engage in auditing and punish misbehaviors or malicious behaviors in the environment. 
Our evaluation shows that \texttt{TAB} is efficient in the simulated Ethereum, and achieves the security, privacy and trustworthiness goals.

\section*{Acknowledgment}
This work was performed while James Joshi was serving as a Program Director at NSF; and the work represents the views of the authors and not that of NSF.
Chao Li is supported in part by Beijing Natural Science Foundation (No. M22039) and National Key R\&D Program of China (No. 2020YFB2103802).

\bibliographystyle{IEEEtran}
\bibliography{references}



\end{document}